\pdfoutput=1
\documentclass[prd,twocolumn,showpacs,showkeys,superscriptaddress,citeautoscript,nofootinbib]{revtex4-1}

\usepackage{amsmath,amssymb}

\usepackage{hyperref}
\hypersetup{pdfauthor={Simeon Hellerman and Shunsuke Maeda and Domenico Orlando and Susanne Reffert and Masataka Watanabe},pdftitle={S-duality and correlation functions at large R-charge},%
  colorlinks=true}

\usepackage{physics}

\newcommand*{\del}{\partial}
\newcommand*{\Op}{\mathcal{O}}
\usepackage{xcolor}
\definecolor{Purple}{rgb}{0.808594, 0.242188, 0.46875}
\newcommand*{\JJM}{{\color{purple}\mathcal{J}}}

\RenewDocumentCommand\ordersymbol{}{O} %
\RenewDocumentCommand\order{ l m }{\fbraces#1{\lparen}{\rparen}{\ordersymbol}{#2}} %

\usepackage{acronym}

\usepackage{graphicx}
\graphicspath{{plots/}}

\usepackage[T1]{fontenc}
\usepackage[utf8]{inputenc}

\usepackage{microtype}

\begin{document}

\title{S-duality and correlation functions at large R-charge}

\author{Simeon Hellerman}
\affiliation{Kavli Institute for the Physics and Mathematics of the Universe. The University of Tokyo.  Kashiwa, Chiba  277-8582, Japan}
\author{Shunsuke Maeda}
\affiliation{Kavli Institute for the Physics and Mathematics of the Universe. The University of Tokyo.  Kashiwa, Chiba  277-8582, Japan}
\author{Domenico Orlando}
\affiliation{\textsc{infn} sezione di Torino, via Pietro Giuria 1, 10125 Torino, Italy}
\affiliation{Albert Einstein Center for Fundamental Physics, Institute for Theoretical Physics, University of Bern, Sidlerstrasse 5, 3012 Bern, Switzerland}
\author{Susanne Reffert}
\affiliation{Albert Einstein Center for Fundamental Physics, Institute for Theoretical Physics, University of Bern, Sidlerstrasse 5, 3012 Bern, Switzerland}
\author{Masataka Watanabe}
\affiliation{Department of Particle Physics and Astrophysics, Weizmann Institute of Science, Rehovot 7610001, Israel}

\begin{abstract}
  We study the ratio of pairs of adjacent correlators of Coulomb-branch operators in \(SU(2)\) \(\mathcal{N}=2\) \acs{sqcd} with four flavors within the framework of the \acl{lqne}.
  Capitalizing on the order-by-order S-duality invariance of the large-R-charge expansion, we compute \emph{ab initio} the dependence of the leading large-$\JJM$ behavior on the marginal coupling \(\tau\) and we find excellent agreement with numerical estimates from localization. %
\end{abstract}

\maketitle

\acrodef{lqne}[\textsc{lqne}]{large quantum number expansion}
\acrodef{vev}{vacuum expectation value}
\acrodef{qft}[\textsc{cft}]{quantum field theory}
\acrodefplural{qft}[\textsc{qft}s]{quantum field theories}
\acrodef{cft}[\textsc{cft}]{conformal field theory}
\acrodefplural{cft}[\textsc{cft}s]{conformal field theories}
\acrodef{scft}[\textsc{scft}]{superconformal field theory}
\acrodef{eft}[\textsc{eft}]{effective field theory}
\acrodefplural{eft}[\textsc{eft}s]{effective field theories}
\acrodef{sqcd}[\textsc{sqcd}]{supersymmetric QCD}
\acrodef{ir}[\textsc{ir}]{infrared}
\acrodef{uv}[\textsc{uv}]{ultraviolet}
\acrodef{agt}[\textsc{agt}]{Alday--Gaiotto--Tachikawa}

\acused{sqcd}
\acused{agt}

\section{Introduction}%
\label{sec:intro}

The \ac{lqne}~\cite{Hellerman:2015nra,Hellerman:2017sur,Hellerman:2018xpi, Alvarez-Gaume:2016vff, Monin:2016jmo} is a surprisingly precise approximation scheme that applies to generic \acp{qft} with global symmetries and allows to compute correlation functions in strongly coupled systems.
While generally, not much is known about such systems, in the case of
theories with extended superconformal symmetry, it is possible to check the predictions of the \ac{lqne} against supersymmetric localization results.

In theories with a gauge group of rank one, the correlation functions of Coulomb branch operators of large R-charge \(\JJM\) can be computed in terms of the \ac{eft} of a single $U(1)$ vector multiplet, with a two-derivative kinetic term, and a single higher-derivative term proportional
to the $a$-anomaly mismatch between the $U(1)$ \ac{eft} and the underlying \ac{scft}~\cite{Hellerman:2017sur,Hellerman:2018xpi}.
The case of
one-dimensional Coulomb branch is particularly tractable because there are no superconformal $F$-terms that can be constructed for a single $U(1)$ vector
multiplet.
In general, the only ambiguities of observables in an \ac{eft} are unknown Wilson coefficients plus corrections exponentially small in the ratio of the \ac{uv} to the \ac{ir} energy scale.
Since the F-term sector of the $U(1)$ \ac{eft} contains no terms with adjustable coefficients other than the kinetic term and Euler term, the correlation functions
are computable to all orders in the inverse $R$-charge, modulo those two unknown coefficients.
Specifically, the correlation function of two Coulomb branch operators with scaling dimension $\Delta$ is given by the product of a theory-dependent factor \(C_n\) and a universal one:
\begin{equation}
\label{eq:two-point-function}
  \ev*{\Op^n_\Delta(x)\bar{\Op}^n_\Delta(y)} = C_n \frac{\Gamma(\JJM + \alpha + 1)}{\abs{x -y}^{2\JJM}} ,
\end{equation}
where \(\alpha\) measures the $a$-anomaly mismatch between the underlying \ac{cft} and the moduli-space \ac{eft}\footnote{The normalization of the $a$-anomaly is chosen so that a free $U(1)$ vector multiplet has $a = 5/24$, and $\alpha \equiv 2a$ in those units.}, $\JJM \equiv n\Delta$  and the $n$-dependence in \(C_n\) takes the form
\begin{equation}
  C_n = e^{n A + \tilde B} + \order{e^{-\kappa\sqrt{n}}} ,
\end{equation}
where $\kappa$ is a coefficient depending on the theory and on the marginal coupling if any.
The $A$ and $\tilde{B}$ coefficients depend on the (scheme-dependent) kinetic term and Euler term in the $U(1)$ \ac{eft}, and
the normalization of the chiral ring generator $\Op_\Delta$.

In checking the predictions of the large-charge \ac{eft} against other methods such as supersymmetric localization,
one must either check combinations of correlators in which the $A$ and $\tilde{B}$ coefficients cancel (as done in the literature so far~\cite{Hellerman:2017sur,Hellerman:2018xpi,Grassi:2019txd}), or then carefully match the scheme-dependent $A$ and $\tilde{B}$ coefficients between the large-charge \ac{eft} and the microscopic \ac{cft}.
While possible in principle, this is somewhat challenging for general $\mathcal{N} = 2$ \acp{scft} because in general there is no preferred or canonical normalization
for the chiral ring generator, so any scheme choice would contain an element of arbitrariness.
In theories with a marginal coupling, on the other hand, there do exist a preferred class of normalizations for the chiral ring generator, associated with the tangent bundle over the ``conformal manifold'' (the conformally invariant subspace of theory space).  For rank-one
theories with marginal coupling, the chiral ring is generated by a single operator \(\Op\) whose normalization is
related to the normalization of a holomorphic tangent vector to theory space.
More specifically, for theories with an $S$-duality
symmetry, the chiral ring generator transforms in a specific way under $S$-duality. In practical terms this property gives enough information to fix a
preferred normalization for the chiral ring generator, and gives a preferred scheme in which to compute the dependence of the $A$-coefficient on the marginal coupling.

In this note we consider the case of $\mathcal{N} = 2$ \(SU(2)\) \ac{sqcd} with \(N_f = 4\), using its $S$-duality symmetry to match the normalization of the
chiral ring generator between the large-charge \ac{eft} and the underlying \ac{cft}.
With this matching specified, the $A$-coefficient is well-defined as a function of the marginal coupling, and we can use the consistency conditions of the \ac{eft} together with the supersymmetric recursion relations~\cite{Papadodimas:2009eu,Baggio:2014ioa,Baggio:2014sna,Baggio:2015vxa,Gerchkovitz:2016gxx} to
calculate it.
In terms of the marginal coupling $\tau$ with respect to which
the recursion relations are satisfied, we find that (up to exponentially-suppressed gauge-instanton corrections)
\begin{equation}
  A(\tau, \bar \tau) = \log(\frac{1}{4(2 \Im(\tau) + 4/ \pi \log(2))^2} )\ .
\end{equation}
With the $A$-coefficient specified, we can calculate the large-charge expansion of ratios of \emph{pairs of adjacent correlators}, with the result
\begin{equation}
  \frac{\ev*{\Op^n(x) \bar{\Op}^n(y)}}{\ev*{\Op^{n-1}(x) \bar{\Op}^{n-1}(y)}} \simeq \frac{(4n + 1) (4n + 3 ) }{64 \pqty{ \Im(\tau) +  \frac{\log(4)}{\pi} }^2 \abs{x -y }^{4}},
\end{equation}
where the $\simeq$ indicates corrections going as $e^{- \kappa \sqrt{n}}$ at fixed $\tau$, and as $e^{2\pi i \tau}$ at fixed $n$.
Checking this against supersymmetric localization results, we find highly precise numerical agreement of the order of one part per ten thousand or better in the appropriate regime.

The outline of this note is as follows: first we recall some results in the \ac{lqne} for rank-one \acp{scft} and show that the \(A\) coefficient obeys a Liouville equation (Section~\ref{sec:recursion-relations}).
Then we discuss the relationship between two important parametrizations of the conformal manifold, in terms of the microscopic coupling \(\tau\) and the S-duality-covariant coupling \(\sigma\) (Section~\ref{sec:scheme-dependence}).
In the \(\sigma\)-frame the correlation functions transform as nonholomorphic modular forms under the action of S-duality (Section~\ref{sec:operators-as-modular-forms}), which imposes strong constraints on the form of the solution to the Liouville equation (Section~\ref{sec:Liouville-sigma}).
The solution is completely identified by matching with the known weak-coupling asymptotic behavior (Section~\ref{sec:unique-solution}).
Finally, we compare our prediction with numerical results from localization (Section~\ref{sec:numerics}).

\section{Recursion relations and large $R$-charge}%
\label{sec:recursion-relations}

The strategy used in~\cite{Hellerman:2017sur,Hellerman:2018xpi} is
to combine some elementary properties of the large-charge \ac{eft} with
the supersymmetric recursion relations~\cite{Papadodimas:2009eu,Gerchkovitz:2016gxx} obeyed
by correlation functions of Coulomb-branch chiral primaries in gauge theories
with complex 
marginal coupling $\tau$.  The recursion relations are expressed most
naturally in terms of the quantities
\begin{equation}
  G_{2n}= \ev*{\Op^n(x) \bar{\Op}^n(y)} \abs{x - y}^{4n}
\end{equation}
or their
logarithms $q_n = \log( Z_{S^4} G_{2n})$.  With respect to $\tau$, the
recursion relations take the form of a semi-infinite Toda lattice equation~\cite{Papadodimas:2009eu,Baggio:2014ioa,Baggio:2014sna,Baggio:2015vxa,Gerchkovitz:2016gxx}:
\begin{equation}
  \label{eq:Toda}
  \del_\tau \del_{\bar \tau} q_n = e^{q_{n+1}-q_n} - e^{q_n - q_{n-1}} .
\end{equation}
At large $n$, the system is driven into an Abelian Coulomb phase, as 
the operator insertion creates a source for the Coulomb branch modulus,
which acquires a \ac{vev} of order $n^{1/2}$. For $n$ larger than $g_{\textsc{ym}}^{-2}$, the off-diagonal gauge
bosons can be integrated out, and the Coulomb-branch \ac{eft}
becomes parametrically accurate at finite gauge coupling.

The \ac{eft} picture sets a boundary condition for
the recursion relations at large $n$~\cite{Hellerman:2017sur}, with the
result that the asymptotics must be
\begin{equation}
  Z_{S^4} G_{2n} = (2n)! \, e^{n A} n^{\alpha} \bqty{ \order{n^0} + \order{n^1} + \dots },
\end{equation}
where $\alpha$ is fixed by a Weyl $a$-anomaly coefficient, and the coefficient $A$ depends on the theory.  These 
asymptotics hold regardless of whether or not the theory has marginal
parameters; if the theory does have marginal parameters, then the $A$ coefficient
depends on those as well.
In this case, the relation in Eq.~\eqref{eq:Toda} implies that \(A(\tau, \bar \tau)\) is a solution to the Liouville equation~\cite{Hellerman:2018xpi}
\begin{equation}
  \del_\tau \del_{\bar \tau} A(\tau, \bar \tau) = 8 e^{A(\tau, \bar \tau)} .
\end{equation}

Without reference to the recursion relations, the $A$ coefficient depends
on the additional input of the normalization of the chiral ring generator
$\Op$ itself.  For a theory with marginal parameters obeying the
recursion relations, though, this datum is fixed automatically~\cite{Papadodimas:2009eu,Gerchkovitz:2016gxx} by the relationship between the
marginal coupling and the Coulomb branch chiral primary.
The latter is a superconformal descendant of the former, which gives a specific normalization in the relationship between the marginal operator and the
chiral primary.  

This relationship, however, does not resolve the normalization issue \emph{per se}, because a marginal operator itself has no preferred normalization either; rather, a marginal operator is normalized naturally as an element of the tangent bundle over the conformal manifold.
That is, a marginal operator \emph{transforms as a tangent vector} under reparametrization of the coupling constant $\tau$.
As a result, the chiral ring generator $\Op(\tau)$ transforms with a holomorphic
Jacobian under change of scheme \(\tau \to \tau' = f(\tau)\):
\begin{align}
  \Op(\tau) \to \Op'(\tau') = \eval{\frac{1}{f'(\tau)} \Op(\tau) }_{\tau = f^{-1}(\tau')}.
\end{align}

\section{Scheme-dependence and duality in $\mathcal{N} = 2$ superconformal SQCD}%
\label{sec:scheme-dependence}

In our problem there are two preferred frames for the parametrization of the conformal manifold:
\begin{itemize}
\item The \(\tau\)-frame, in which the recursion relations are derived. This is the same normalization scheme
for the marginal operator in which the $S^4$ partition function
was calculated by Pestun in~\cite{Pestun:2007rz}.  
This scheme also has a manifest crossing symmetry when interpreted
{via} two-dimensional quantum Liouville theory under the \ac{agt} correspondence~\cite{Alday:2009aq}.
\item The \(\sigma\)-frame, in which $S$-duality symmetry is realized with the modular group acting on the coupling \(\sigma\) by rational functions.
\end{itemize}
The two frames coincide in the case of \(\mathcal{N}=4\) super-Yang--Mills, but they are different for the conformal \ac{sqcd} with $G = SU(2)$ and $N_F = 4$.
In the \(\tau\)-frame the $S$-duality symmetry acts as a change of
channel in a four-point function of local operators under the \ac{agt} correspondence.%

Based on this, it is possible to give a map between the microscopic coupling $\tau$
and the S-duality-covariant coupling \(\sigma\) (called \(2\tau_{\textsc{ir}}\) in~\cite{Alday:2009aq})
in terms of the \emph{modular lambda function} $\lambda(\sigma)$~\cite{Grimm:2007tm}:
\begin{equation}
  e^{2 \pi i \tau} = \lambda(\sigma).
\end{equation}
To make the relationship between the couplings more concrete we define
\begin{align}
  q &= e^{2\pi i \tau} , & y &=e^{\pi i \sigma} .
\end{align}
Expanding the modular lambda function at large $\Im(\sigma)$, we have
\begin{equation}
  \lambda(\sigma) \simeq 16 y - 128 y^2 + 704 y^3 - 3072 y^4 + 11488 y^5 + \dots
\end{equation}
so that the relationship between the couplings is
\begin{align}
  \label{SeriesExpansionForSigmaAsAFunctionFOfTauPaperSummary}
  \sigma &\equiv F(\tau) = 2 \tau +  \frac{4 i}{\pi} \log(2) - \frac{i}{\pi} \pqty{\frac{q}{2} + \frac{13}{64} q^2 + \frac{23}{192} q^3 + \dots   } \\
  \tau &\equiv L(\sigma) = \frac{\sigma}{2} - \frac{2i}{\pi} \log(2) + \frac{i}{\pi} \pqty{4 y - 6 y^2+ \frac{16}{3} y^3 + \dots } .
\end{align}
For most of the discussion of the large-R-charge expansion, the instanton corrections will be too small to be relevant.%
The constant additive contribution can be thought of as the threshold correction to the effective Abelian
gauge coupling contributed by integrating out the massive $W$-bosons and hyper multiplets, with
the $\log(2)$ expressing the logarithmic running of the Abelian gauge coupling between
the mass of the W-bosons and the mass of the hypers, which differ by a factor of $2$.  This
$\mathcal{N} = 2$ threshold correction played an important role historically in Seiberg--Witten theory; 
the nonzero value of the correction was the first clue indicating a distinction between
the microscopic coupling $\tau$ and the naively S-duality-covariant coupling $\sigma$, which 
were in effect assumed to be equal in the early literature~\cite{Seiberg:1994aj}.
We will see that this threshold correction is distinctly visible to high numerical accuracy as a term
in the large $R$-charge expansion of the Coulomb branch correlation functions, compared
across various different values of the microscopic gauge coupling $\tau$.

In the following we will use a hatted notation to distinguish the operators in the \(\sigma\) frame as opposed to the \(\tau \) frame.
Let $\Op(\tau )$ be the chiral ring generator, \emph{i.e.} the operator whose insertion at the pole is equivalent to differentiating \(\log(Z_{S^4})\) by $\tau$~\cite{Gerchkovitz:2016gxx}. Instead of a ``microscopic'' definition of \(\hat \Op\) we will give an operational definition: it is the object whose insertion is defined by differentiating the logarithm of the partition function with
respect to the \emph{modular-covariant} coupling\footnote{As always, $g_{\textsc{ym}}^2$ is defined so that the single gauge instanton has action $S_{\text{single instanton}} = 8 \pi^2/g_{\textsc{ym}}^2$ in the weak-coupling limit.} $\sigma = 8\pi i/g_{\textsc{ym}}^2 + \theta/\pi$.
The two are related by
\begin{equation}
  \hat \Op(\sigma) = \eval{ \dv{\tau}{\sigma} \Op(\tau)}_{\tau = L(\sigma)} = L'(\sigma) \Op(L(\sigma)) . %
\end{equation}
Given the transformation law for the operators, we come to the conclusion that when we change the frame from \(\tau\) to \(\sigma\), the correlators 
\begin{equation}
  G_{2n}(\tau, \bar \tau) = \ev*{(\Op^n(x) \bar \Op^n(y)} \abs{x - y}^{2 n \Delta_{\Op}}
\end{equation}
transform as
\begin{equation}
  G_{2n}(\tau, \bar \tau) \to \hat G_{2n}(\sigma, \bar \sigma) = \abs{L'(\sigma)}^{2n} G_{2n}(L(\sigma),\overline{L(\sigma)}). %
\end{equation}

\section{Chiral primary operators as holomorphic modular forms}%
\label{sec:operators-as-modular-forms}

How does the correlation function of the Coulomb branch operator \(\hat \Op\) in the \(\sigma\)-frame transform under modular transformations?
First recall that a modular transformation is a  Möbius map of the upper half-plane of the type
\begin{align}
  \sigma \to \sigma' &= \frac{\alpha \sigma + \beta}{\gamma \sigma + \delta}, &
                                      \begin{pmatrix}
                                        \alpha & \beta \\ \gamma & \delta
                                      \end{pmatrix} \in SL(2, \mathbb{Z}).
\end{align}
A function \(E(\sigma, \bar \sigma)\) is a modular form of weight \((k,l)\) if under a modular map it transforms as
\begin{equation}
  E(\sigma, \bar \sigma) \to E(\sigma', \bar \sigma') = (\gamma \sigma + \delta)^{k} (\gamma \bar \sigma + \delta)^{l} E(\sigma, \bar \sigma).
\end{equation}

If we differentiate the logarithm of the sphere partition function twice, once each with respect to $\sigma$ and
$\bar \sigma$, we get an object whose modular transformation has to be the same as that of $a^2 \bar a^2$, where
$a$ is the vacuum modulus that enters the central charge $Z$ with coupling-independent coefficient
\begin{equation}
  Z =  a ( n_e + \sigma n_m) .
\end{equation}
Under the $S$-transformation $\sigma \to - 1/\sigma$, $a$ transforms as $a\to \sigma a$, and under the $T$ transformation
$\sigma \to \sigma + 1$, $a$ is invariant.  It follows that \(a \) transforms as a \((1,0)\) modular form
\begin{equation}
  a \to ( \gamma \sigma + \delta) a .
\end{equation}
But then, the correlation function for the operator \(\hat \Op \) \emph{must transform as a nonholomorphic modular form} of weight $(2,2)$:
\begin{equation}
  \ev*{\hat\Op(\sigma ) \hat {\bar\Op}( \bar \sigma )} \to (\gamma \sigma + \delta)^{2} (\gamma \bar \sigma + \delta)^{2} \ev*{\hat\Op(\sigma ) \hat {\bar\Op}( \bar \sigma )} . 
\end{equation}
More in general, the correlation function \(\hat G_{2n}\) is a modular form of weight \((2n,2n)\), and \(e^{\hat A}\) is a modular form of weight \((2,2)\).

\section{Liouville equation in the $\sigma$-frame}
\label{sec:Liouville-sigma}

In the \(\tau\)-frame, the Toda lattice equation implies that the function \(A(\tau, \bar \tau)\) satisfies the Liouville equation~\cite{Hellerman:2018xpi}
\begin{equation}
  \label{eq:Liouville-eqn-tau-frame}
  \del_\tau \del_{\bar \tau} A(\tau, \bar \tau) = 8 e^{A(\tau, \bar \tau)} .
\end{equation}
What happens in the change of frame?
The Liouville equation is actually conformally covariant if we let the Liouville field transform affinely with the log of the Jacobian, \emph{i.e.} if we let $e^A$ transform like a complex differential form of bidegree \((1,1)\):
\begin{align}
  e^{\hat A(\sigma, \bar \sigma)} = \abs{\dv{\tau}{\sigma}}^2 e^{A(\tau, \bar \tau)}.
\end{align}
Then \(\hat A(\sigma, \bar \sigma)\) obeys the same Liouville equation:
\begin{equation}
  \label{LiouvilleEquationWithSigmaFrameExplicitlyIndicated}
  \del_\sigma \del_{\bar \sigma} \hat A(\sigma, \bar \sigma) = 8 e^{\hat A(\sigma, \bar \sigma)} .
\end{equation}
In a simply-connected domain the general solution is written in terms of an arbitrary holomorphic function \(f(\sigma )\) and an arbitrary antiholomorphic function   \(\tilde f(\bar \sigma)\):
\begin{equation}
  \label{eq:Liouville-general-solution}
  e^{\hat A(\sigma, \bar \sigma)} = \frac{\del_\sigma f(\sigma) \del_{\bar \sigma} \tilde f(\bar \sigma)}{( 1 - 4 f(\sigma) \tilde f(\bar \sigma))^2} .
\end{equation}

In the \(\sigma\)-frame we know that the quantity $e^{\hat A (\sigma, \bar \sigma )}$ must transform
as a modular form of weight $(2,2)$; this sets a useful boundary condition for
the Liouville equation.  In practical terms, this boundary condition is most easily used by imposing a boundary condition on a derived holomorphic quantity $T(\sigma) $ which we will construct below, that transforms under $S$-duality as a holomorphic modular form of weight $4$.
This is unique (up to an overall constant to be fixed matching with weak-coupling results), which will give us sufficient information to constrain the solution $\hat A(\sigma, \bar \sigma )$.

Starting from the general form of the solution to the Liouville equation in Eq.~\eqref{eq:Liouville-general-solution}, we take the Schwarzian derivatives of \(f(\sigma)\) and \(\tilde f(\bar \sigma)\),\footnote{These are essentially the holomorphic stress tensors of classical Liouville theory, \emph{cf. e.g.}~\cite{Seiberg:1990eb}.} \(T(\sigma)\) and \(\tilde T (\bar \sigma)\):
\begin{align}
  T(\sigma) &= \frac{f'''(\sigma)}{f'(\sigma)} - \frac{3}{2} \pqty{\frac{f''(\sigma)}{f'(\sigma)} }^2 = \del_{\sigma}^2 \hat A(\sigma, \bar \sigma) - \frac{1}{2} (\del_{\sigma} \hat A(\sigma, \bar \sigma))^2, \\
  \tilde T(\bar \sigma) &= \frac{\tilde f'''(\bar \sigma)}{\tilde f'(\bar \sigma)} - \frac{3}{2} \pqty{\frac{\tilde f''(\bar \sigma)}{\tilde f'(\bar \sigma)} }^2 = \del_{\bar \sigma}^2 \hat A(\sigma, \bar \sigma) - \frac{1}{2} (\del_{\bar \sigma} \hat A(\sigma, \bar \sigma))^2 .
\end{align}
Using the fact that \(e^{\hat A}\) is a modular form of weight \((2,2)\), we have that under modular transformation \(T(\sigma)\) transforms as
\begin{equation}
  T(\sigma ) \to T(\sigma') = (\gamma \sigma + \delta)^4 T(\sigma),
\end{equation}
and is a modular form of weight \((4,0)\).
In fact, this property fixes \(T(\sigma)\) up to an overall constant.
By construction \(T(\sigma)\) is holomorphic as a function of \(\sigma \) (it is the Schwarzian derivative of a holomorphic function) and there is a unique holomorphic modular form of weight \(4\), \emph{i.e.} the Eisenstein series \(E_4(\sigma)\).
Repeating the same reasoning for \(\tilde T\), we find that 
\begin{equation}
  (T(\sigma), \tilde T(\bar \sigma)) = (\xi E_4(\sigma), \tilde \xi E_4(\bar \sigma)) ,
\end{equation}
where \(\xi\) and \(\tilde \xi\) are constants that are to be fixed using the boundary condition supplied by weak-coupling information.

It is important to note that, while \(E_4\) is the unique holomorphic modular form of weight \(4\), there are of course many more meromorphic modular forms of weight $4$, or even more broadly,
locally holomorphic modular forms with branch points, such as for instance $E_6^{2/3}$.  But, since $e^{\hat A}$ is an
\ac{eft} object (an inverse gauge kinetic term-squared), we know it cannot have singularities except at points where the \ac{eft}
breaks down and new particles are integrated in.  This does not happen for any finite coupling in a rank-one \ac{scft}.
The distinction between holomorphic modular forms versus more general locally holomorphic modular forms with the same transformation
property may be relevant when considering more general examples, for instance with higher rank.

\section{Constraints from weak-coupling asymptotics}%
\label{sec:weak-coupling}

Now we state an important fact: The symmetries of the problem, together
with the Liouville equation, %
force the condition $T(\sigma)= \tilde T(\bar \sigma) =  0$.

At weak coupling (\(\Im(\sigma) \to \infty\)), the asymptotics of $E_4(\sigma)$ are
\begin{equation}
  E_4(\sigma) = 1 + 240 e^{2 \pi i \sigma} + \dots
\end{equation}
so, in the weak-coupling region the constants \(\xi\) and \(\tilde \xi\) are
\begin{equation}
  (\xi, \tilde \xi) = \lim_{\Im(\sigma) \to \infty} (T(\sigma), \tilde T(\bar \sigma)) .
\end{equation}

We know from perturbative calculation that at weak coupling~\cite{Bourget:2018obm,Beccaria:2018xxl,Beccaria:2020azj} in the \(\tau\)-frame the correlation function for \ac{sqcd} can be approximated with the one for \(\mathcal{N} = 4\) \(SU(2)\) theory:
\begin{equation}
  G_{2n}(\tau, \bar \tau) =  G_{2n}^{\mathcal{N}=4}(\tau, \bar \tau) \pqty{1 + \order{\Im(\tau)^{-2}}}
\end{equation}
and
\begin{equation}
  G_{2n}^{\mathcal{N}=4}(\tau, \bar \tau) = \frac{\Gamma(2n+2)}{\Im(\tau)^{2n}} .
\end{equation}
In this case the \(\alpha\)  coefficient and the dimension \(\Delta\) in Eq.~\eqref{eq:two-point-function} are respectively \(\alpha = 1\) and \(\Delta = 2\) (see~\cite{Hellerman:2017sur}) so we have that for \(\Im(\tau) \to \infty \),
\begin{equation}
  e^{n A(\tau, \bar \tau)} = \Im(\tau)^{-2n} ( 1 + \order{\Im(\tau)^{-2}}) .
\end{equation}
This last identity takes a bit of explanation.  What we really know about weak coupling, directly, is the two-point function as a function of $\Op_2$ at weak coupling, or more generally the $n$th root
of the two-point
function of $\Op_2^n$ at weak coupling for some fixed $n$.  \emph{A priori} this is not the same thing
as $e^{A(\tau, \bar \tau)}$.  But in fact the two are the same, as can be inferred from the
weak-coupling double-scaling analysis of~\cite{Bourget:2018obm}.
This analysis shows that
the $\tau^{-2n}$ term in the two-point correlator of $\Op_{2n}$ is just the $n$th power of the
$\tau^{-2}$ term in the two-point function of $\Op_2$. %
\begin{widetext}
  We just need to translate to the \(\sigma \) frame:
\begin{equation}
  \hat A(\sigma, \bar \sigma) = A(L(\sigma),\overline{L(\sigma)}) + \log(\abs{L'(\sigma)}^2) %
  = -2 \log(\Im(\sigma)) + \order{\Im(\sigma)^{0}}  ,
\end{equation}
to compute the boundary behavior of \(T\) and \(\tilde T\):
\begin{equation}
  (\xi, \tilde{\xi}) = \lim_{\Im(\sigma) \to \infty } (T(\sigma), \tilde T(\bar \sigma)) 
    = (\del_\sigma^2 \hat A(\sigma, \bar \sigma) - \frac{1}{2}(\del_\sigma \hat A(\sigma, \bar \sigma))^2, \del_{\bar \sigma}^2 \hat A(\sigma, \bar \sigma) - \frac{1}{2}(\del_{\bar \sigma} \hat A(\sigma, \bar \sigma))^2  )= (0,0) .
\end{equation}
\end{widetext}

\section{The unique solution}%
\label{sec:unique-solution}

Now that we know that \(T(\sigma) = \tilde T(\bar \sigma) = 0\), we can deduce the unique solution to the Liouville equation that satisfies all the boundary conditions.

The defining property of the Schwarzian derivative is that it is invariant under Möbius transformations.
The derivative  \(T(\sigma) = (\mathcal{S}f)(\sigma )\) vanishes if and only if \(f\) is a rational function of the form
\begin{equation}
  f(\sigma) = \frac{a \sigma + b}{c \sigma + d} .
\end{equation}
If \(f\) is Möbius and \(\hat A\) is real (so that \(\tilde f(\bar \sigma) = \overline{f(\sigma)}\)), a direct calculation shows that \(e^{\hat A}\) has modular weight \((2,2)\) if and only if the coefficients \(a\), \(b\), \(c\), \(d\) satisfy the constraints
\begin{align}
  a \bar d &= b \bar c,  & c \bar d &=4 \bar a b, & d \bar d &=4 b \bar b .
\end{align}
Substituting into the expression of the general solution of Liouville's equation we find
\begin{equation}
  e^{\hat A(\sigma, \bar \sigma)} = \frac{1}{16\, \Im(\sigma)^2}  .
\end{equation}
Transforming back to the $\tau$ coordinate we find
\begin{multline}\label{eq:finalprediction}
  A(\tau, \bar \tau) = \hat A(F(\tau), \overline{F(\tau)}) + \log(\abs{F'(\tau)}^2) \\ = \log(\frac{1}{4(2 \Im(\tau) + 4/\pi \log(2) )^2} ) + \order{e^{2 \pi i \tau}} .
\end{multline}
This prediction is our main result.
It agrees extremely well with the numerical results from localization in a large range of couplings and values of $n$ as we will see in the next section.

\section{Numerics from localization}%
\label{sec:numerics}

In the previous section we have derived an expression for the ratio of successive correlation functions, that was all-orders in $n$ with an exponentially small error, and \emph{exact} as a function of the gauge coupling $\tau$.  In this section we will compare this expression with exact results from localization.

In the \(\tau\)-frame, the function \(G_{2n}(\tau, \bar \tau )\) can be written in terms of the ratio of determinants of normalized matrices of derivatives of the partition function \(Z_{S^4}\)~\cite{Gerchkovitz:2016gxx}:
\begin{align}
  \eval{\mathcal{M}_{(n)}}_{kl} &= \frac{1}{Z_{S^4}} \del^k_{\tau} \del^l_{\bar \tau} Z_{S^4}, && k,l = 1, \dots, n-1 \\
  G_{2n}(\tau, \bar \tau) &=   \frac{\abs{\mathcal{M}_{(n)}}}{\abs{\mathcal{M}_{(n-1)}}} .
\end{align}
In turn, the partition function is an integral of Barnes \(G\)-functions (which are analytic continuations of the superfactorial):
\begin{gather}
  Z_{S^4} = \int \dd{a} a^2 e^{-4a \Im(\tau)} \frac{\abs{G(1 + 2 i a)}^4}{\abs{G(1 + i a)}^{16}} \abs{Z_{\text{inst}}(a, \tau)} ^2, \\
  G(n) = (n-2)\$ = \prod_{k=1}^{n-2}k!,
\end{gather}
where \(Z_{\text{inst}}\) is the instanton partition function
\begin{equation}
  Z_{\text{inst}}(a,\tau) = 1 + \frac{1}{2} \pqty{a^2 - 3} e^{2 \pi i \tau} + \order{e^{4\pi i \tau}}.
\end{equation}
In the regime \(\Im(\tau) > 1\) we can neglect the effect of the instantons since already at \(\Im(\tau) =2\) we find \(e^{2 \pi i \tau} = \order{10^{-6}}\) and evaluate the integral numerically.
We have done so for integer values of \(n\) between \(1 \) and \(35\) and of \(\Im(\tau)\) between \(1 \) and \(50\).
This gives us a way to verify our predictions.
Once more we stress that the comparison of the linear term in \(n\) between the \ac{lqne} and localization is only meaningful when the two computations are performed in the same scheme \emph{i.e.} with respect 
to the same definition of the holomorphic gauge coupling.

In Figure~\ref{fig:DeltaQ-prediction} we show the values of the first variation \(\Delta_n q_n(\tau, \bar \tau) = q_{n+1}(\tau, \bar \tau) - q_n(\tau, \bar \tau)\) as function of \(n\) at fixed values of \(\Im(\tau)\) and as function of \(\Im(\tau)\) at fixed values of \(n\).
The agreement is excellent, as one can verify quantitatively by fitting the numerical data to the functional form \(A(\tau, \bar \tau) = -k_3 \log((2 \Im(\tau) + k_2)/k_4)\).
The best fit of the parameters \(k_i\) is for \(k_2 = 0.8844(4)\), \(k_3 = 2.00067(6)\), \(k_4 = 3.992(1)\), to be compared to the \ac{lqne} predictions of \(k_2 = 4 \log(2)/\pi \approx 0.8825\dots\), \(k_3 = 2\), \(k_4 = 4\).

This agreement between the \ac{lqne} prediction and the numerical data allows us to go beyond the perturbative large-charge expansion that is resummed in the expression in Eq.~\eqref{eq:two-point-function} and discuss the contributions associated to the breakdown of the \ac{eft}.
The natural expectation is to see exponential corrections associated to massive states with masses above the \ac{ir} scale.
These corrections are most easily isolated in a double-scaling limit in which one considers simultaneously large charge, \(n \to \infty\), and weak coupling, \(\Im(\tau) \to \infty \), keeping the ratio \(\lambda = 4 \pi n / \Im(\tau)\) fixed~\cite{Bourget:2018obm,Beccaria:2018xxl}.
In this limit the gauge-instanton corrections to the partition function \(Z_{S^4}(\tau, \bar \tau)\) are negligible, and we can meaningfully compare the \ac{lqne} and the localization computation.
The final result is that the difference is extremely well approximated by an exponential of the type
\begin{equation}
  \eval{q_n}_{\ac{lqne}} - \eval{q_n}_{\text{loc}} \simeq k_5 e^{-k_6 \sqrt{n/\Im(\tau)}}
\end{equation}
with  \(k_5 = 1.842(6)\) and \(k_6 = 3.347(13)\) (see Figure~\ref{fig:final-error-prediction-exponential}).
Using this fit we find an agreement of the order of one part per ten thousand at \(\lambda = 8 \pi\) and one part per million at \(\lambda = 32 \pi\).%

\section{On the exponentially small correction}
This type of exponentially small correction associated to the propagation of particles with mass of order \(\order{\lambda^{1/2}}\) has also been found in the double-scaling limit of~\cite{Grassi:2019txd}. The setup is however different since, while the \ac{lqne} is S-duality-invariant term-by-term in the expansion, the double-scaling limit is not S-duality covariant.
For this reason the two expansions will generically differ.

More specifically, it is not clear that the exponent $k_6$ need agree with the exponent of the corresponding exponential correction computed in the douple-scaling limit of the full correlator.  Although both corrections
have a double-scaling form and a physical interpretation as a massive particle worldline on the scale set by $|x-y|$, there is no clear sense in which one is a limit of the other.

The exponent
of the exponential correction estimated 
here and in~\cite{Hellerman:2018xpi} comes from subtracting the universal all-orders formula from the full localization result, and taking a double-scaling limit of the difference.  By contrast the exponent of the exponential
correction in~\cite{Grassi:2019txd}, is obtained from the double-scaling limit of the full correlator.
Since the universal fixed-coupling \ac{eft} factor does not itself have a double-scaling limit, the relationship between the two
exponential corrections is unclear theoretically; there is no \emph{a priori} reason known to us to expect them to be equal.
Numerically they are close, but the difference may be significant and could shed light on the relationship
between the fixed-coupling large-charge behavior and the double-scaling large-charge behavior.  
Computing the exponent directly by a worldline computation -- either in the double-scaling limit of the microscopic theory or
in the \ac{eft} of the fixed-coupling large-charge limit or preferably in both limits -- would be a valuable piece of data to clarify the issue and we are hopeful this can be attempted in future work.

\section{Conclusions}%
\label{sec:conclusions}

The \ac{lqne} approach is based on the use of symmetries.
In this work we have seen how to take advantage of the S-duality symmetry to discuss the \(\tau\)-dependence of the two-point function of Coulomb branch operators.

Here we have discussed the behavior of the correlation functions of Coulomb branch operators in \(\mathcal{N}=2\) \(SU(2)\) \ac{sqcd} with \(N_f = 4\).
Specifically, we have computed the dependence of the ratio of pairs of adjacent correlators as a function of the marginal coupling \(\tau\).
In the framework of the \acl{lqne}, this ratio depends on a function \(A(\tau, \bar \tau)\) that is a solution to the Liouville equation.
The crucial observation is that in an appropriate parametrization of the conformal manifold, the quantity \(e^A\) transforms (after rescaling by a Jacobian corresponding to a change of scheme) as a nonholomorphic modular form of weight \((2,2)\) under the S-duality symmetry of the theory.
This requirement, together with the matching of the known weak-coupling behavior of the correlation functions, fixes the function \(A\) completely.
We have also compared the predicted behavior with numerical estimates obtained using localization.
This required a careful matching of the scheme-dependent quantities is made possible by the observation that the generator of the Coulomb branch transforms as an element of the tangent bundle over the conformal manifold.
The agreement is strikingly good and has allowed us to estimate the leading behavior of the exponential corrections that signal the breakdown of the \ac{eft}.

While here we have used S-duality invariance to constrain some correlators, we expect to be able to push this approach further and use it to derive stringent conditions on the spectrum of this theory beyond the perturbative regime of the \ac{lqne}.
The \ac{lqne} must be S-duality-invariant order-by-order in $n$.  All the power-law terms satisfy this trivially, being independent of $\tau$ identically.
But, the exponentially small corrections are constrained by S-duality in a useful way, since their behavior depends nontrivially  on the gauge coupling.
The constraints of S-duality on the exponentially small corrections will be perhaps most concretely realized by the modular bootstrap equation around fixed points of finite-order $SL(2, \mathbb{Z})$ transformations of the conformal manifold. This Letter forms the starting point of the study of S-duality at large R-charge.

\section*{Ackowledgments}%

S.H. thanks the authors of~\cite{Grassi:2019txd} for sharing their numerical data and for useful discussions on the relationship between the exponentially small corrections
in the double-scaling limit and those in the fixed-coupling large-$\JJM$ limit.
D.O. would like to thank M.Billò and F.Galvagno for insightful comments and discussions.  
The work of S.H. is supported by the World Premier International Research
Center Initiative (\textsc{wpi} Initiative), \textsc{mext}, Japan; by the \textsc{jsps} Program for Advancing Strategic International Networks to Accelerate the Circulation of Talented Researchers;
and also supported in part by \textsc{jsps kakenhi} Grant Numbers \textsc{jp22740153, jp26400242}.
D.O. acknowledges partial support by the \textsc{nccr 51nf40--141869} ``The Mathematics of Physics'' (Swiss\textsc{map}).
The work of S.R. is supported by the Swiss National Science Foundation under grant number \textsc{pp00p2\_183718/1}.  The authors also thank the Simons Center for Geometry and Physics for hospitality during the program, ``Quantum Mechanical Systems at Large Quantum Number,'' during which this work was initiated.

\begin{widetext}

  \begin{figure}[b]
  \begin{tabular}{lr}
    \includegraphics[width=.49\textwidth]{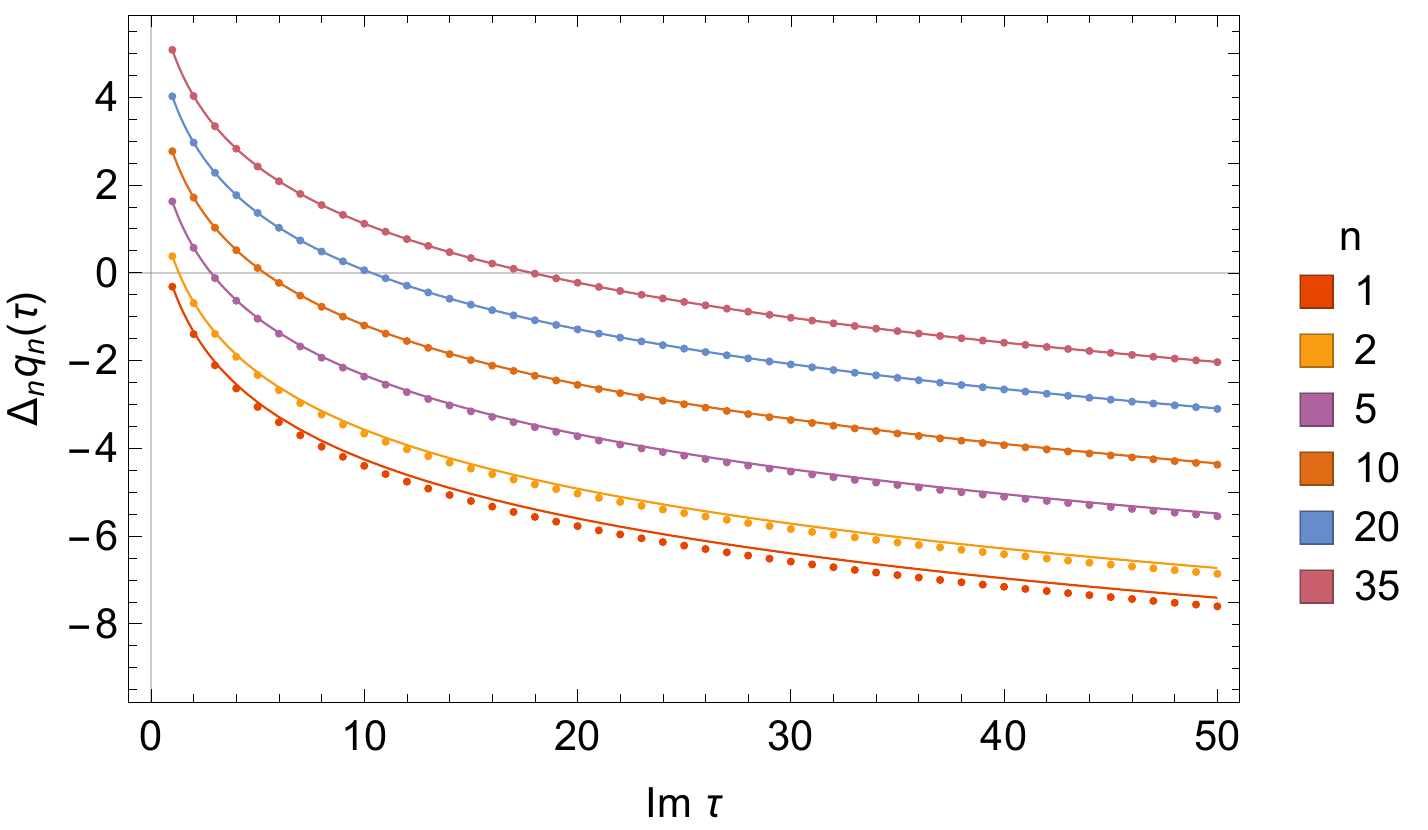} &
                                                                 \includegraphics[width=.49\textwidth]{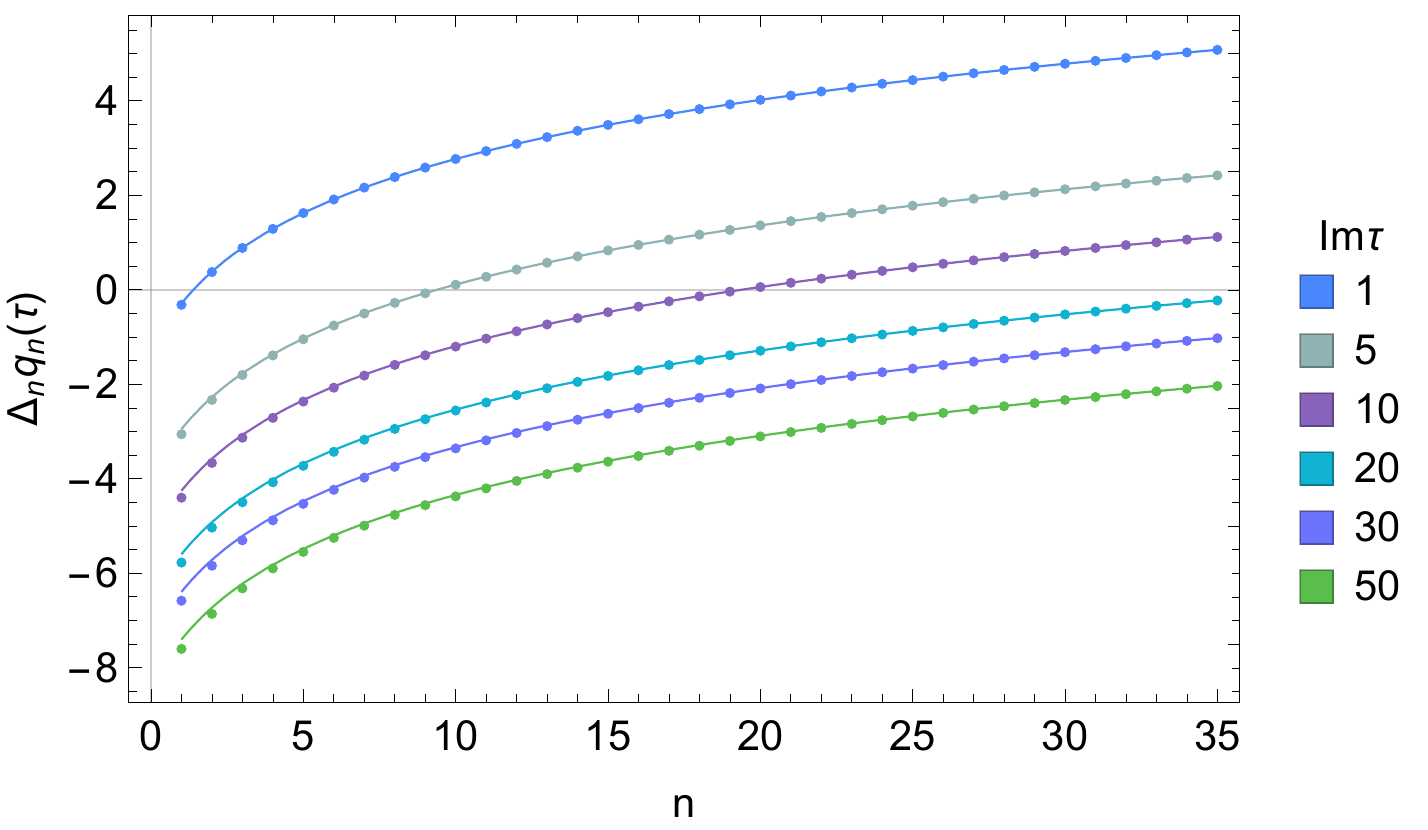}
  \end{tabular}
  \caption{First variation \(\Delta_n q_n(\tau)\) as function of \(\Im(\tau)\) for fixed values of \(n\) (left) and as function of \(n\) for fixed values of \(\Im(\tau)\) (right). The dots are numerical values estimated from localization, the continuous lines are the \ac{lqne} prediction in Eq.~\eqref{eq:finalprediction}.}
  \label{fig:DeltaQ-prediction}
\end{figure}

\begin{figure}
    \begin{tabular}{lr}
      \includegraphics[width=.49\textwidth]{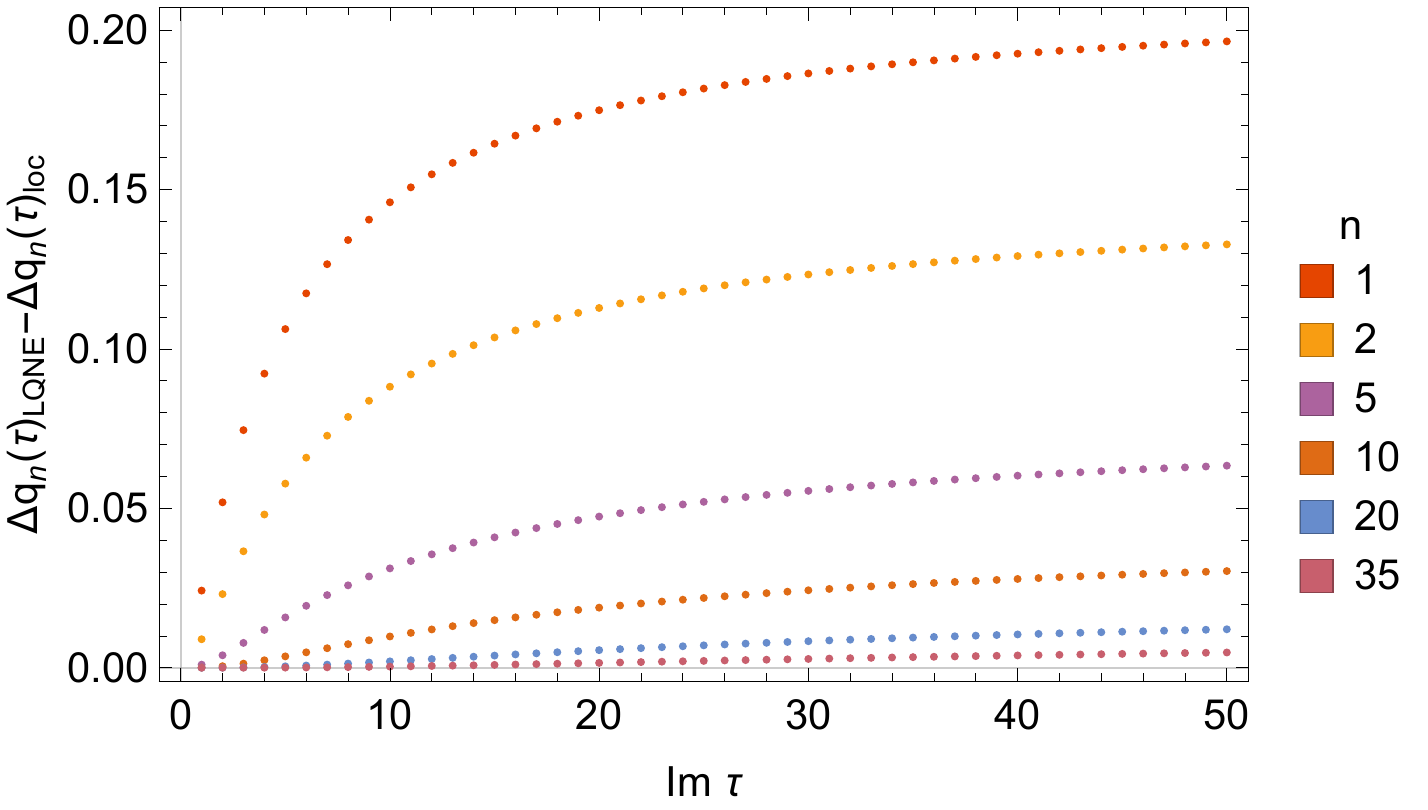} &
                                                                  \includegraphics[width=.49\textwidth]{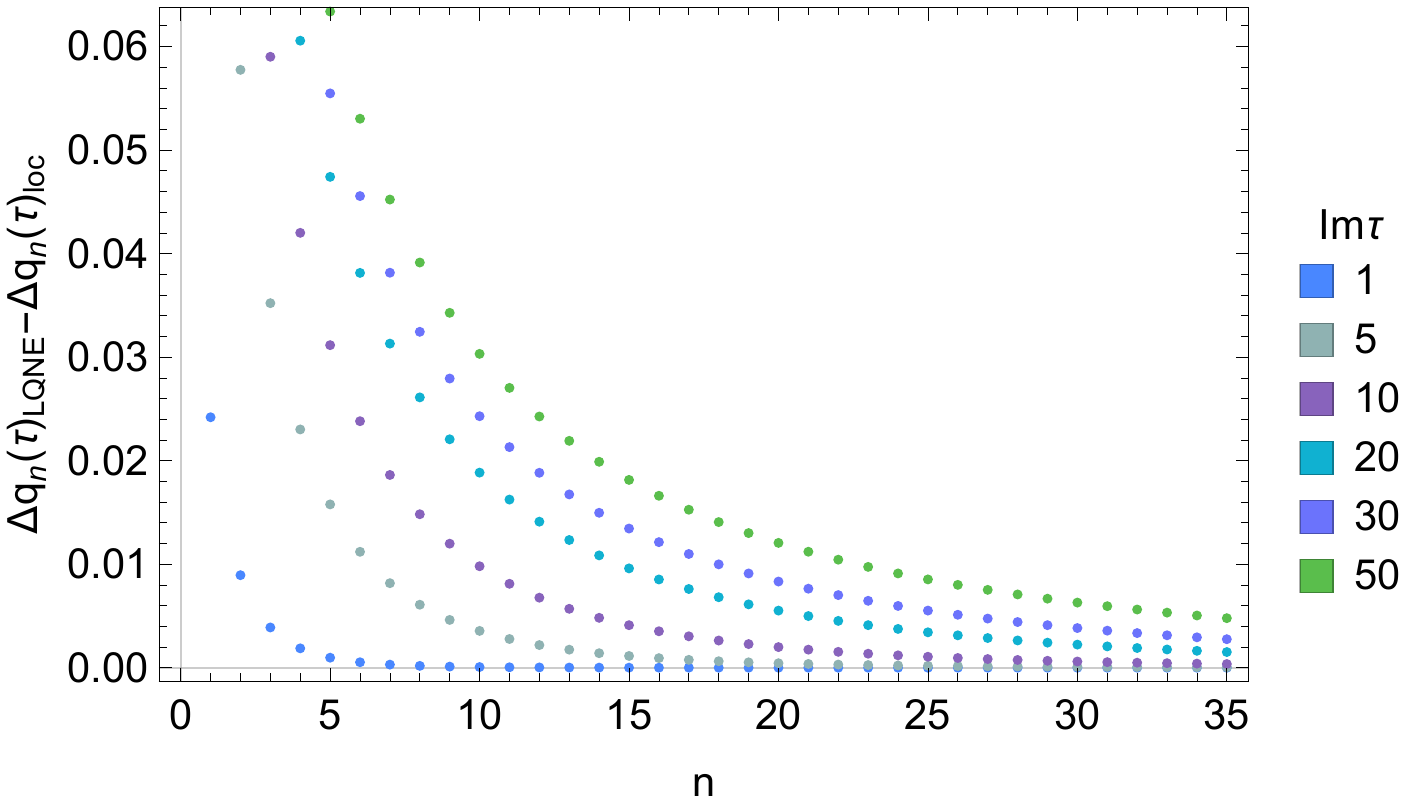} %
    \end{tabular}
  \caption{Difference between the \ac{lqne} prediction and the localization estimate for the first variation \(\Delta_n q_n\) as function of \(\Im(\tau)\) for fixed values of \(n\) (left) and as function of \(n\) for fixed values of \(\Im(\tau)\) (right). %
  Note that the agreement between the \ac{lqne} and the exact result is best at strong coupling and worst (though still extremely precise) at weak coupling, as expected from the interpretation of the
  exponentially small correction as attributable to the breakdown of the \ac{eft} from the macroscopic propagation of a massive hypermultiplet, whose mass goes as the square root of the gauge coupling.
  }
  \label{fig:error-prediction}
\end{figure}

\begin{figure}
  \centering
  \includegraphics[width=.49\textwidth]{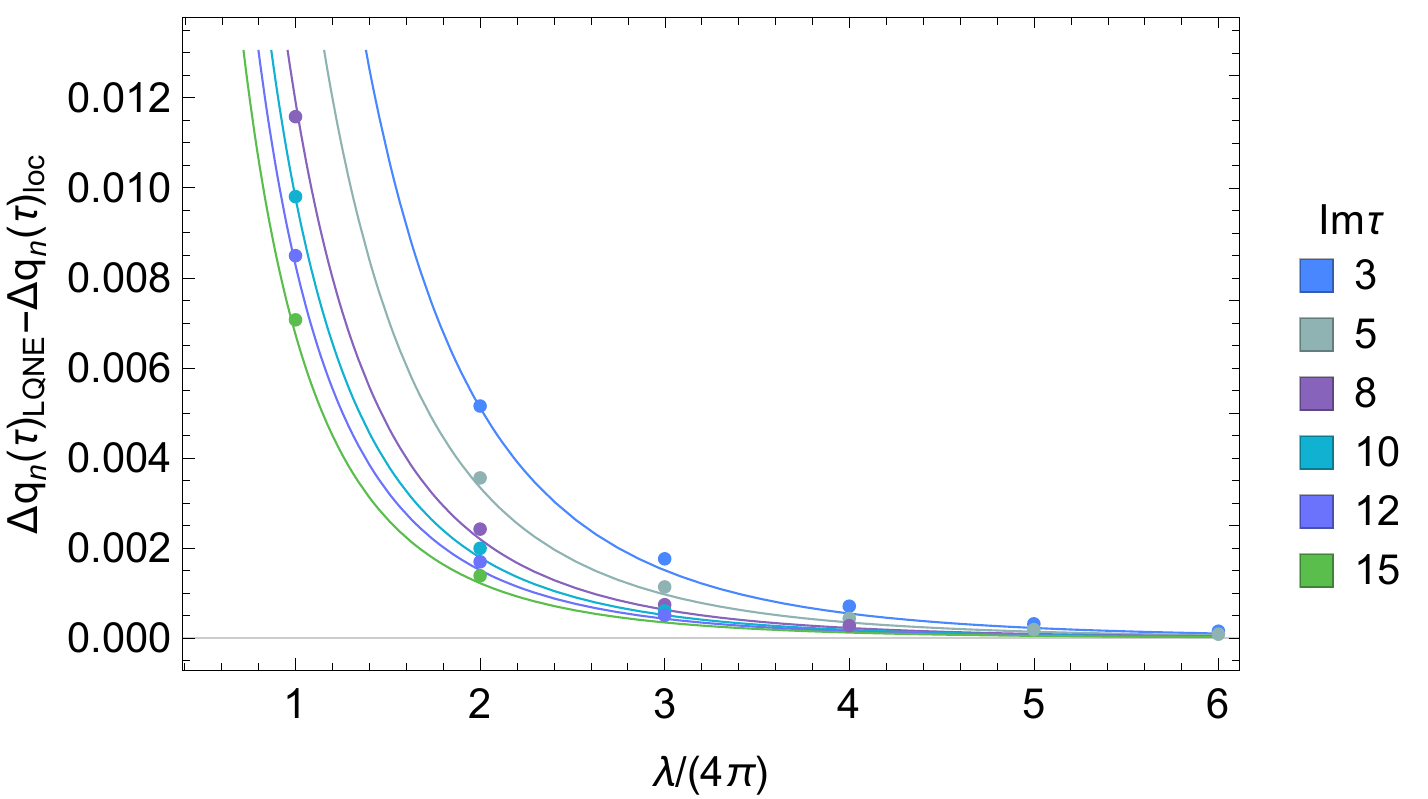}
  \caption{Difference between the \ac{lqne} prediction and the localization estimate for the first variation \(\Delta_n q_n\) as function of \(\lambda/(4\pi) = n/\Im(\tau)\) for fixed values of \(\Im(\tau)\) (dots). The continuous lines are the best numerical fit for the exponential correction \(k_5 \exp(-k_6 \sqrt{n/\Im(\tau)})\) with  \(k_5 = 1.842(6)\) and \(k_6 = 3.347(13)\).}
  \label{fig:final-error-prediction-exponential}
\end{figure}

\end{widetext}

\bibliography{references}

\end{document}